\documentstyle[11pt,epsfig]{article}

\newcommand{\be}{\begin{equation}}
\newcommand{\ee}{\end{equation}}
\newcommand{\ba}{\begin{array}}
\newcommand{\ea}{\end{array}}
\newcommand{\bea}{\begin{eqnarray}}
\newcommand{\eea}{\end{eqnarray}}
\newcommand{\bean}{\begin{eqnarray*}}
\newcommand{\eean}{\end{eqnarray*}}
\newcommand{\eqn}[1]{(\ref{#1})}

\newcommand{\dy} {\displaystyle}

\def\ru1{\rule[-0.4truecm]{0mm}{1truecm}}
\newcommand{\gapproxeq}{\lower .7ex\hbox{$\;\stackrel{\textstyle >}{\sim}\;$}}
\newcommand{\lapproxeq}{\lower .7ex\hbox{$\;\stackrel{\textstyle <}{\sim}\;$}}
\def\upa{\uparrow}
\def\dna{\downarrow}
\def\nn{\nonumber}
\def\sbar{{\bar s}}
\def\ubar{{\bar u}}
\def\dbar{{\bar d}}
\def\qbar{{\bar q}}
\def\dde{\Delta}
\def\al{\alpha}
\def\bet{\beta}

\def\de{\delta}

\def\th{\Theta}

\def\ccfr{{\scriptscriptstyle CCFR}}
\def\nc{{\scriptscriptstyle NC}}
\def\cc{{\scriptscriptstyle CC}}

\def\nlo{{\scriptscriptstyle NLO}}
\def\nmc{{\scriptscriptstyle NMC}}
\def\als{{\alpha_s}}
\def\alpiq{\frac{\als (Q^2)}{\pi}}
\def\q0{Q_0^2}

\def\vud2{|V_{\scriptscriptstyle ud}|^2}
\def\vus2{|V_{\scriptscriptstyle us}|^2}
\def\vcd2{|V_{\scriptscriptstyle cd}|^2}
\def\vcs2{|V_{\scriptscriptstyle cs}|^2}

\def\up#1{\leavevmode \raise.16ex\hbox{#1}}

\def\sqr#1#2{{\vcenter{\vbox{\hrule height.#2pt
	\hbox{\vrule width.#2pt height#1pt \kern#1pt
	  \vrule width.#2pt}
	\hrule height.#2pt}}}}

\newcommand{\jou}[4]{{\rm #1 }{\bf #2} \up(19#3\up) #4}

\newcounter{appendix}

\def\thebibliography#1{{\bf REFERENCES\markboth
 {REFERENCES}{REFERENCES}}\list
 {[\arabic{enumi}]}{\settowidth\labelwidth{[#1]}\leftmargin\labelwidth
 \advance\leftmargin\labelsep
 \usecounter{enumi}}
 \def\newblock{\hskip .11em plus .33em minus -.07em}
 \sloppy
 \sfcode`\.=1000\relax}

\begin{document}

\thispagestyle{empty}

{\hfill hep-ph/0001159}

{\hfill DSF-42/99}\vspace*{1cm}

\begin{center}
{\Large \bf The Strange Quark Problem in the Framework}

{\Large \bf of Statistical Distributions}
\end{center}

\bigskip\bigskip

\begin{center}
{\bf F. Buccella, O. Pisanti, and L. Rosa}
\end{center}

\begin{center}
\begin{tabular}{l}
Dipartimento di Scienze Fisiche, Universit\`a di Napoli, 
Mostra d'Oltremare, Pad.19, \\
~~~I-80125, Napoli, Italy; \\
INFN, Sezione di Napoli, Napoli, Italy.
\end{tabular}
\end{center}

\begin{abstract}
A large class of polarized and unpolarized deep inelastic data is
successfully described with Fermi-Dirac functions for the non-diffractive
part of quark parton distributions. The NLO approach used here improves the
agreement with experiment of the previous LO work. We get a broader
distribution for the strange parton $s(x)$ than for $\sbar(x)$. 
\end{abstract}

\bigskip


\newpage

\section{Introduction}

The possibility of having $s(x) \neq \sbar(x)$ has been advocated
\cite{bro96} to explain the conflict between two different determinations
of the strange quark sea in the nucleon. The CTEQ global analysis
\cite{cteq93} obtained the strange quark distribution on the assumption of
$s(x)=\sbar(x)$. Actually, the measured quantity was the following
combination of nucleon structure functions on an isoscalar target, $N$: 
\be
\frac{5}{12}~ \left( F_2^{\nu N} + F_2^{\bar \nu N} \right) - 3~ F_2^{\mu
N} = \frac{x}{2}~ [s(x) + \sbar(x)], 
\ee
where the sum of $F_2$ neutrino structure function came from CCFR DIS data
from neutrino and antineutrino beams \cite{ccfrold} and $F_2^{\mu N}$ from
the NM Collaboration \cite{nmcold}. The result of this mean
strange-antistrange quark distribution was found to be quite different from
the strange quark distribution extracted from dimuon events of CCFR
\cite{ccfrdimu1}. Since the neutrino events dominate the CCFR data set, one
should consider the latter result as $s(x)$. Therefore one may consider the
conflict as a first evidence for the difference between $s(x)$ and
$\sbar(x)$. Note that as a result of the reanalysis by CCFR
\cite{ccfrdimu2}, including the effect of higher-order QCD corrections, the
observed discrepancy is reduced, but it is still significant. An
interesting contribution to this problem would be represented by an
independent evidence of $s(x) \neq \sbar(x)$. New neutrino and antineutrino
data on $F_2$ and $F_3$ have been measured \cite{ccfrnew} and they are
quantities (especially $F_3$) sensitive to $s(x) \neq \sbar(x)$. Therefore
a global analysis of nucleon structure functions, including $F_2$ and $F_3$
from neutrino and antineutrino beams rather than the dimuon events of CCFR
measurement, may provide this kind of independent confirmation. 

On the theoretical side, with the heuristic argument that $m_\Lambda/3 >
m_K/2$ \cite{bro96} a broader distribution for $s(x)$ (expected to combine
with the valence $u$ and $d$ quarks to give a $\Lambda$) than for
$\sbar(x)$ (expected to combine with the same quarks to give a $K$) was
advocated. This is just the property that could explain the previously
mentioned data conflict. 

In another scenario, a different shape for $s(x)$ and $\sbar(x)$ may be
naturally understood in the framework of an approach developed in the last
years \cite{pauli,posi,bhal}. With the motivation of keeping into account
the role played by Pauli principle in explaining several experimental
facts, the parton distributions were described \cite{posi} in terms of the
sum of a {\it gas} component, given by a Fermi-Dirac function for quarks
($i=u^\upa$, $u^\dna$, $d^\upa$, $d^\dna$, $\ubar$, $\dbar$), 
\be
p_{i,gas} (x) = \frac{A~ x^\al~ (1-x)^\bet}{\exp\left(\frac{x-\tilde
x_i}{\bar x}\right) + 1}, 
\label{e:qpauliold}
\ee
and a {\it liquid} term, unpolarized and isoscalar, given by
\be
L (x) = 0.12~ x^{-1.19}~ (1-x)^{9.6}.
\ee
For the gluons a Bose-Einstein function was taken ($i=g^\upa$, $g^\dna$),
\be
p_{i,gas} (x) = \frac{8}{3}~ \frac{A~ x^\al~ (1-x)^\bet}{\exp \left(
\frac{x-\tilde x_i}{\bar x}\right) - 1}. 
\label{e:gpauliold}
\ee
In Eq.s~\eqn{e:qpauliold} and \eqn{e:gpauliold} $\bar x$ plays the role of
a {\it temperature}, $\tilde x_i$ of a {\it thermodynamic potential},
depending on the flavour and the spin of each parton, and the function in
the numerator is a weight function for the density levels of the quarks in
the $P_z=\infty$ frame of reference. In Ref. \cite{posi} this statistical
parametrization was applied for describing both polarized and unpolarized
data and compared with a standard one, obtaining a very good agreement
between the first, second, and third moment of the distributions. 

However, it is rather natural to assume different weight functions for $q$
and $\qbar$, since in the nucleon, which is not a $C$ invariant object, the
quarks (transforming as a 3 under $SU(3)_c$) dominate over the antiquark
(transforming as a $\bar 3$). Indeed, inspired by the previously mentioned
heuristic argument ($m_\Lambda/3 > m_K/2$), we expect a broader weight
function for the quarks than for the antiquarks. The strange quark is the
best one to test this idea, since it is the lightest not valence parton. 

With respect to the analysis of Ref. \cite{posi}, that was performed at
Leading Order (LO) in the strong coupling, $\al_s$, here we include the
Next to Leading Order (NLO) corrections. We study a large class of
polarized and unpolarized data within the framework just described, with
the important modification of taking different weight functions for quarks
and antiquarks. 

The paper is organized as follows. In Section 2 the input parametrizations
of the parton distributions are presented, Section 3 is devoted to the
description of the experimental data used in our analysis and of the method
of resolution of evolution equations, and Section 4 reports our results and
conclusions. 

\section{Parton Parametrizations}

According to the hypothesis of a role of Pauli principle for quark parton
distributions, we take, at $Q_0^2 = 3\, GeV^2$, Fermi-Dirac functions for
$u$ and $d$ quarks ($i=u^\upa$, $u^\dna$, $d^\upa$, $d^\dna$), 
\be
p_i (x,Q_0^2) = \frac{f_q(x)}{\exp\left(\frac{x-\tilde x_i}{\bar x}\right)
+ 1} + \frac{1}{2}~ f_L (x), 
\label{e:qpauli}
\ee
where the last term,
\be
f_L (x) = A_L~ x^{\al_L}~ (1-x)^{\bet_L}, 
\ee
represents an unpolarized and isoscalar part which takes into account the
large diffractive contribution in the small $x$ region ($-2< \al_L\leq
-1$). 

With the aim of allowing different shapes for $s(x)$ and $\sbar(x)$, we
take a different form for the weight function of antiquarks. Since at small
$x$ the single parton contribution is overwhelmed by the diffractive one,
it is very difficult to determine the difference in the exponents of $x$.
So, we assume them equal for quarks and antiquarks, allowing, instead,
different values for the exponents of $(1-x)$, which is sensitive to the
large $x$ behaviour where the diffractive contribution becomes negligible.
In conclusion we have 
\bea
f_q (x) &=& A_q~ x^\al~ (1-x)^{\bet_q} \label{e:fq}, \\
f_\qbar (x) &=& A_\qbar~ x^\al~ (1-x)^{\bet_\qbar}, \label{e:fqbar}
\eea
with $\al>-1$.

In the analysis of Ref. \cite{posi} we found rather negative values for the
{\it potentials} of the partons in the sea, which correspond to the
Boltzmann limit for their distributions. We then parametrize $s$, $\ubar$,
and $\dbar$ in the following way: 
\bea
\ubar(x,Q_0^2) &=& k_\ubar~ f_\qbar(x)~ e^{-\frac{x}{\bar x}} + f_L (x), \\
\dbar(x,Q_0^2) &=& k_\dbar~ f_\qbar(x)~ e^{-\frac{x}{\bar x}} + f_L (x), \\
s(x,Q_0^2) &=& k_s~ f_q(x)~ e^{-\frac{x}{\bar x}} + \frac{2}{4.2}~ f_L (x).
\label{e:s}
\eea
Note that the different coefficient in front of $f_L (x)$ in Eq.~\eqn{e:s}
takes into account the results of Ref. \cite{ccfrdimu2} in the relative
size of the strange sea with respect to the non-strange one. This result is
also used to decrease the number of parameters, in writing 
\be
\sbar (x,Q_0^2) = \frac{\ubar(x,Q_0^2) + \dbar(x,Q_0^2)}{4.2},
\ee
with, however, the requirement that the first moments of s and $\sbar$ be
equal, since the nucleons have no strangeness. 

The defect in the Gottfried sum rule \cite{gottfr} implies a non trivial
sea in the nucleons, with $\dbar > \ubar$, as confirmed also by the
experiments NA51 at CERN \cite{na51} and E866 at FNAL \cite{e866}. Along
this line, we should also consider the possibility of polarization in the
quark sea, as advocated to account for the large defect in the Ellis and
Jaffe sum rule \cite{elljaf} for the polarized structure function of the
proton, $g_1^p$, first shown in the EMC experiment \cite{emc} and confirmed
by the following ones. We expect the distributions to be proportional to
the {\it gas} component of the unpolarized ones, 
\bea
\dde \ubar(x,Q_0^2) &=& \dde k_\ubar~ [\ubar(x,Q_0^2) - f_L (x)], \\ 
\dde \dbar(x,Q_0^2) &=& \dde k_\dbar~ [\dbar(x,Q_0^2) - f_L (x)], \\
\dde s(x,Q_0^2) + \dde \sbar(x,Q_0^2) &=& \dde k_{s \sbar}~ \left[
s(x,Q_0^2) + \sbar(x,Q_0^2) - \frac{4}{4.2} f_L (x) \right], 
\eea
with $|\dde k_i| \leq 1$ and for simplicity we have taken the same
proportionality constant for $s$ and $\sbar$. The isovector combination
$\dde \ubar - \dde \dbar$ is thus proportional to the {\it gas} component
of $\ubar + \dbar$ by a factor, which is less than one in modulus for the
positivity constraints on polarized distributions, and reaches that value
only in the extreme cases of full and opposite polarization for $\ubar$ and
$\dbar$. For the isoscalar combination we have a part proportional to
$f_\qbar$, $\dde \ubar + \dde \dbar$ and $\dde \sbar$, and another one,
$\dde s$, proportional to $f_q$. In conclusion, we have three parameters,
$\dde k_\ubar$, $\dde k_\dbar$, and $\dde k_{s \sbar}$ to describe the sea
contribution to the two polarized structure functions of the nucleons,
$g_1^p$ and $g_1^n$. 

For the gluon distribution we take the Bose-Einstein form ($i=g^\upa$,
$g^\dna$), 
\be
p_i (x,Q_0^2) = \frac{f_g (x)}{\exp\left(\frac{x-\tilde x_i}{\bar x}\right)
- 1}, 
\label{e:gpauli}
\ee
with $f_g$ given by
\be
f_g (x) = \frac{8}{3}~ A_g~ x^{\al_g}~ (1-x)^{\bet_g},
\ee
but with the constraint $\al_g \leq \al$. Note that a more divergent term,
similar to the second one in the r.h.s. of Eq.~\eqn{e:qpauli}, seems to be
excluded by data; so, we chose not to include it in Eq.~\eqn{e:gpauli}. 

We impose that the total momentum of the partons be unity. Other
constraints on the distributions concern the values of the following
unpolarized sum rules: 
\bea
I_{Adler} &=& 1.01 \pm 0.20, \label{e:adl} \\
I_{GLS} &=& 2.5 \pm 0.08, \label{e:gls} \\
I_{Gottfried} &=& 0.235 \pm 0.026. \label{e:gott} 
\eea
As far as the Adler sum rule \cite{adler} is concerned, we consider this
constraint as an exact one, while we use the experimental errors for the
other ones. Moreover, in calculating the theoretical value of the
Gross-Llewellin-Smith sum rule \cite{gls} we include the QCD corrections up
to $O\left( \al_s^3\right)$ \cite{glscorr}, 
\be
I_{GLS} (Q^2) = I_{GLS}^{(0)} (Q^2) \left[ 1 - \alpiq - 3.25 \left( \alpiq
\right)^2 - 12.20 \left( \alpiq \right)^3 \right], 
\ee
where
\be
I_{GLS}^{(0)} (Q^2) = \int_0^1~ dx [u-\ubar+d-\dbar+s-\sbar] (x,Q^2).
\ee
Finally, we constrain the ratio $(\ubar/\dbar) (x=0.18)$ to the value
$0.51\pm 0.06$ measured by the experiment NA51 \cite{na51}. 

\section{Description of Data and Parton Evolution}

We perform a NLO description, in the $\overline{MS}$ scheme, of the
unpolarized data on $F_2^\nu$ and $F_3^\nu$ from CCFR \cite{ccfrnew},
$F_2^{p,d}$ from NMC \cite{nmcnew}, and of the polarized structure
functions measured at SLAC in the E142 \cite{e142}, E143 \cite{e143}, and
E154 \cite{e154} experiments, at HERA in the HERMES experiment
\cite{hermes}, and at CERN in the SMC experiment \cite{smc}. All the
unpolarized data were subjected to the cuts $Q^2 \geq 2\, GeV^2$ and $W^2
\geq 10\, GeV^2$. We add a $1.5\%$ uncertainty to the statistical errors of
CCFR data since no global systematic errors are given. 

For solving the DGLAP evolution equations \cite{dglap} we use the Jacobi
polynomial method \cite{jac}, which some of us already used in an analysis
of polarized structure functions \cite{evjac}. Here, we briefly recall the
procedure. The given structure function is expressed in terms of a
truncated series of Jacobi polynomials, $\th_k^{\al\bet}(x)$,
\be
F(x,Q^2) = x^\al (1-x)^\bet~ \sum_{k=0}^N~ a_k^{(\al\bet)} (Q^2)
\th_k^{\al\bet}(x), 
\ee
where $\th_k^{\al\bet}$ satisfy the orthogonality condition
\be
\int_0^1~ x^\al (1-x)^\bet~ \th_k^{\al\bet} \th_l^{\al\bet}~ dx = \de_{kl}.
\ee
By using the definition of the coefficients, $a_k^{(\al\bet)}$,
\be
a_k^{(\al\bet)}(Q^2) = \int_0^1~ F(x,Q^2)~ \th_k^{\al\bet} (x)~ dx,
\ee
with a little algebra it is possible to obtain for $F(x,Q^2)$ the following
expression, 
\be
F(x,Q^2) = x^\al (1-x)^\bet~ \sum_{k=0}^N~ \th_k^{\al\bet}(x)~
\sum_{j=0}^k~ c_j^{(k)} (\al,\bet)~ F_{j+1} (Q^2), 
\label{e:fjac}
\ee
where $c_j^{(k)} (\al,\bet)$ are the coefficients of the expansion of the
Jacobi polynomials, $\th_k^{\al\bet}(x)$, in power of $x$, and $F_n (Q^2)$
are the Mellin moments of $F(x,Q^2)$, 
\be
F_n(Q^2) = \int_0^1 x^{n-1}~ F(x,Q^2)~ dx.
\ee
In this way, the $Q^2$ dependence of $F(x,Q^2)$ is factorized in its
moments, for which the solution of the evolution equations up to NLO is
well known. 

In this analysis we reconstructed the unpolarized structure functions with
$N=12$ and the polarized ones with $N=9$. Moreover, we used different
values of $\al$ and $\bet$ for the different data sets. Table 1 reports the
values of the parameters which give the best convergence of the Jacobi
expansion. 

The unpolarized parton distributions at $Q_0^2$ are combined to give the
following non-singlet ($Q_i$), singlet ($\Sigma$) and gluon terms ($G$) (we
suppress for brevity the $x$ and $Q_0^2$ dependence of the various terms), 
\bea
Q_p &=& 2~ (u + \ubar) - (d + \dbar) - (s + \sbar), \label{e:firstcomb} \\
Q_n &=& 2~ (d + \dbar) - (u + \ubar) - (s + \sbar), \\ 
Q_3 &=& u - \ubar + d - \dbar + s - \sbar, \\ 
Q_s &=& s - \sbar, \\
\Sigma &=& u + \ubar + d + \dbar + s + \sbar, \\ 
G &=& g,
\eea
while in the polarized case we have 
\bea
\de Q_p &=& a_3 + \frac{a_8}{3} = \frac{4}{3}~ (\dde u + \dde \ubar) -
\frac{2}{3} (\dde d + \dde \dbar + \dde s + \dde \sbar), \\ 
\de Q_n &=& - a_3 + \frac{a_8}{3} = \frac{4}{3}~ (\dde d + \dde \dbar) -
\frac{2}{3} (\dde u + \dde \ubar + \dde s + \dde \sbar), \\ 
\de \Sigma &=& a_0 = \dde u + \dde \ubar + \dde d + \dde \dbar + \dde s +
\dde \sbar, \\ 
\de G &=& \dde g. \label{e:lastcomb} 
\eea

The moments of the previous quantities at $Q_0^2$ are evolved to the $Q^2$
of data (see \cite{evjac} for the complete expressions of the evolved
moments). The relation between the evolved moments of the combinations in
Eq.s~\eqn{e:firstcomb}-\eqn{e:lastcomb} and the structure function moments,
to be used in Eq.~\eqn{e:fjac}, comes from the expression of the
unpolarized and polarized structure functions. Neglecting for the moment
the charm contribution, we have for the first ones (for neutrino structure
functions we give the expressions for isoscalar targets) 
\bea
\frac{F_2^{ep} (x,Q^2)}{x} &=& \left\{ C_2^q \otimes \left( \frac{1}{9}~
Q_p + \frac{2}{9}~ \Sigma \right) \right\} (x,Q^2) + (C_2^g \otimes G)
(x,Q^2), \\ 
\frac{F_2^{en} (x,Q^2)}{x} &=& \left\{ C_2^q \otimes \left( \frac{1}{9}~
Q_n + \frac{2}{9}~ \Sigma \right) \right\} (x,Q^2) + (C_2^g \otimes G)
(x,Q^2), \\ 
F_2^{ed} (x,Q^2) &=& \frac{1}{2}~ (F_2^{ep} (x,Q^2) + F_2^{en} (x,Q^2)), \\
\frac{F_2^\ccfr (x,Q^2)}{x} &=& \frac{1}{x}~ \left[ F_2^{(\nu + \bar\nu)} +
(2~ \al - 1)~ F_2^{(\nu - \bar\nu)} \right] = \left \{ C_2^q \otimes \left[
\frac{1+\vud2}{6}~ (2~ \Sigma + Q_p + Q_n) + \right. \right. \nn \\ 
&& \!\!\!\!\!\!\!\!\!\!\!\!\!\!\!\!\!\!\!\!\!\!\!\!\!\!\!\!\!\!
\frac{\vus2}{3} \left. \left. (\Sigma - Q_p - Q_n) + (2~ \al - 1)~ \left(
\frac{-\vus2}{2}~ (Q_3 - Q_s) + \vus2~ Q_s \right) \right] \right\} (x,Q^2)
+ \nn \\ 
&& \!\!\!\!\!\!\!\!\!\!\!\!\!\!\!\!\!\!\!\!\!\!\!\!\!\!\!\!\!\! 
(C_2^g \otimes G) (x,Q^2), \\ 
F_3^\ccfr (x,Q^2) &=& F_3^{(\nu + \bar\nu)} = \left\{ C_3^q \otimes \left[
\frac{1+\vud2}{2}~ (Q_3 - Q_s) + \vus2~ Q_s \right] \right\} (x,Q^2) 
\label{e:xf3}.
\eea
In the previous equations $C_i^q$ and $C_2^g$ are the coefficient functions
for quarks and gluons, which at LO are given by 
\be
C_i^q (x) = \de (1-x), \quad\quad\quad\quad C_2^g (x) = 0,
\ee
and at NLO can be found, for example, in \cite{buras}. The convolution
$\otimes$ is defined as 
\be
(f \otimes g) (x) \equiv \int_x^1~ \frac{dz}{z} f \left( \frac{x}{z}
\right)~ g(z) = \int_x^1~ \frac{dz}{z} f(z)~ g \left( \frac{x}{z} \right). 
\ee
In the neutrino structure function the Cabibbo-Kobayashi-Maskawa \cite{ckm}
elements appear, for which we use the following values: $\vud2 = 0.9508$,
$\vus2 = 1 - \vud2$, $\vcd2 = 0.0488$, $\vcs2 = 0.9493$. Moreover, $\al =
0.828$ is the fraction of $\nu$ with respect to $\bar\nu$ in the CCFR
experiment \cite{boloth} and a term with $\al$ in the r.h.s. of
Eq.~\eqn{e:xf3} does not appear since the CCFR data have already been
corrected for the $s+\sbar$ contribution \cite{ccfrnew}. The polarized
structure functions are 
\bea
g_1^{p,n} (x) &=& \frac{1}{12}~ \left[ \left( \de C^q \otimes \de Q_{p,n}
\right) (x,Q^2) + \frac{4}{3}~ \left( \de C^g \otimes \de \Sigma \right)
(x,Q^2) \right], \\ 
g_1^d (x) &=& \frac{1}{2}~ \left( 1-\frac{3}{2}~ \omega_D \right)~ \left[
g_1^p (x) + g_1^n (x) \right], 
\eea
where $\de C^i$ are the polarized coefficient functions for quarks and
gluons (see, for example, \cite{nlog1} for their expression) and $\omega_D
= 0.058$ \cite{omd} is the D-wave component in the deuteron ground state.
Note, however, that, when experimentally available, we fit the asymmetries 
\be
A_1^{p,n} (x,Q^2) = \frac{g_1^{p,n} (x,Q^2)}{F_1^{p,n} (x,Q^2)} =
\frac{g_1^{p,n} (x,Q^2)}{F_2^{p,n} (x,Q^2)}~ 2x~ (1+R^{p,n} (x,Q^2)), 
\ee
with $R = F_L/(2\,x\,F_1)$. 

As stressed in \cite{fixed}, a consistent treatment of heavy flavours can
be carried out in the Fixed Flavour Scheme (FFS), where the heavy quarks
are not considered as intrinsic partons, but produced by the interactions
of the other partons (light quarks and gluons). To this aim, we fix the
number of active flavours in the splitting functions to be 3 and add the
charm contributions to neutral and charged current $F_2$ and $x~ F_3$ (we
neglect the small $b$ quark contribution and the $c$ contribution to the
polarized structure function $g_1$), 
\bea
F_{2~ \nc}^{(c)} (x,Q^2) &=& e_c^2~ \frac{\hat{\als}}{\pi}~ x~ \int_{\xi}^1
\frac{dy}{y}~ \hat{C}_2^{g (c)} \left( \frac{x}{y},~ Q^2 \right) \hat{G}
(y), \\ 
F_{2~ \cc}^{(c)} (x,Q^2) &=& 2~ \xi~ \hat{q}' (\xi) +
\frac{\hat{\als}}{\pi}~ \xi~ \sum_{p=q',g}~ \int_{\xi}^1~ \frac{dy}{y}~
\hat{H}_2^p \left( \frac{\xi}{y},~ Q^2 \right) \hat{p} (y),
\label{e:f2nucharm} \\ 
F_3^{(c)} (x,Q^2) &=& 2~ \hat{q}' (\xi) + \frac{\hat{\als}}{\pi}~
\sum_{p=q',g}~ \int_{\xi}^1~ \frac{dy}{y}~ \hat{H}_3^p \left(
\frac{\xi}{y},~ Q^2 \right) \hat{p} (y). \label{e:f3nucharm} 
\eea
In the previous equations the hat indicates that we are calculating a
quantity at $Q^2=\mu^2$ and 
\be
\hat{C} (x,Q^2) \equiv C (x,Q^2,\mu^2), \quad\quad\quad\quad \hat{H}
(x,Q^2) \equiv H (x,Q^2,\mu^2), 
\ee
where $\mu$ is the factorization scale, equal to $4~ m_c^2$ and $Q^2 +
m_c^2$ in the neutral and charged current processes, respectively; the
expressions of the $C$ and $H$'s can be found, for example, in \cite{pgbox}
and \cite{wpbox}; $\xi (x,Q^2) = x~ a(Q^2)$, where 
\be
a(Q^2) = \left\{ 
\ba{ll}
\dy 1 + \frac{4~ m_c^2}{Q^2}, & \quad {\rm for~ NC,} \\
\dy 1 + \frac{m_c^2}{Q^2}, & \quad {\rm for~ CC.}
\ea
\right.
\ee
Furthermore, in Eq.s~\eqn{e:f2nucharm} and \eqn{e:f3nucharm} we have
\be
q' = \left\{
\ba{ll}
\dy \frac{1}{2}~ \left[ \frac{\vcd2}{6}~ (2~ \Sigma + Q_p + Q_n) +
\frac{\vcs2}{3} (\Sigma - Q_p - Q_n)  \right], & \quad {\rm for~ F_2^{\nu +
\bar\nu}}, \\[.5truecm] 
\dy \frac{1}{2}~ \left[ \frac{\vcd2}{2} (Q_3 - Q_s) + \vcs2 Q_s \right], &
\quad {\rm for~ F_2^{\nu - \bar\nu}}, \\[.5truecm] 
\dy \frac{1}{2}~ \left[ \frac{\vcd2}{2} (Q_3 - Q_s) + \vcs2 Q_s \right], &
\quad {\rm for~ F_3}. 
\ea
\right.
\ee

Differently from the charm quark treatment, in the expression of the QCD
coupling constant, we include the usual active flavours, $n_f$, below each
threshold, 
\be
\ba{rcl}
\dy \frac{\al_s^\nlo (Q^2)}{4\, \pi} &=& \dy \frac{1}{\bet_0\,
ln\frac{Q^2}{\Lambda^2}} - \frac{\bet_1}{\bet_0^3}\frac{ln ln
\frac{Q^2}{\Lambda^2}}{ln^2 \frac{Q^2}{\Lambda^2}}, \vspace{.2truecm} \\ 
\bet_0 = 11 - \frac{2}{3} n_f, &\quad\quad& \bet_1 = 102 - \frac{38}{3}
n_f, 
\ea
\ee
where $\Lambda_\nlo^{(5)}$ is fixed to the value 0.2263, so to have $\al_s
(M_Z^2) = 0.118$ \cite{pdg}. Whenever is necessary we use the following
values for the heavy quark masses: 
\be
m_c = 1.5\,GeV,\quad\quad m_b = 4.5\,GeV.
\ee

\section{Results and Conclusions}

We consider the three following cases:

{\bf Fit 1}:
$\dde \ubar$, $\dde \dbar$, $\dde s + \dde \sbar \neq 0$

{\bf Fit 2}:
$\dde \ubar$, $\dde \dbar \neq 0$, $\dde s + \dde \sbar = 0$

{\bf Fit 3}:
$\dde \ubar = \dde \dbar = 0 $, $\dde s + \dde \sbar \neq 0$

In Table 2 the values of the parameters for Fit 1 are reported. We do not
show the results for Fits 2 and 3 because they look very similar to Fit 1
in the parameter values and $\chi_{red}^2$ (1.76 and 1.80, respectively),
with a very small difference in the gluon contribution, which results in a
negligible positive polarization ($\dde g = 0.058$) for Fit 2 and a
slightly larger one ($\dde g = 0.113$) for Fit 3. Table 3 shows the
comparison between the polarization of antiquarks in Fit 1, 2, and 3.
Interestingly enough, $\ubar$ in Fit 2 and $s + \sbar$ in Fit 3 come out
fully negative polarized to confirm that, as supposed in the interpretation
of the EMC result in the gauge-invariant factorization schemes, the sea is
negatively polarized. It is however difficult to disentangle the sea
contribution to the polarization from the valence one, expecially since sea
partons are mostly present at small $x$, where the large diffractive
contribution makes more unprecise the determination of the polarized
structure functions. Indeed, one has to find the polarized cross-section,
which is expected to vanish in the limit $x \rightarrow 0$, from the
difference of two cross-sections, which go to infinity in the same limit.
The values of $a_8 \equiv \dde u + \dde \ubar + \dde d + \dde \dbar - 2
(\dde s + \dde \sbar)$, which are related from $SU(3)$ symmetry to the
combination $3 F - D = 0.579 \pm 0.025$ \cite{fd}, are very different for
Fit 2 (0.373) and 3 (0.884), while Fit 1 is similar to Fit 3 (0.956). This
shows that at the moment we cannot provide a test for this $SU(3)$
prediction. 

In Table 3 the values of the polarized sum rules, Bjorken \cite{bjor},
Ellis-Jaffe \cite{elljaf} for the proton and for the neutron, are also
reported in correspondence of the three considered cases. The values for
the Bjorken sum rule are consistent with the $O(\als^3)$ theoretical
prediction at $Q^2 = 3\,GeV^2$, $0.174\pm 0.002$. 

The quark distributions at $Q_0^2$ are plotted in Fig. 1. In Fig. 2 we
compare $Q_p$, $Q_n$, $\Sigma$, and $g$, evolved to $Q^2 = 1\,GeV^2$, with
the same quantities obtained from the global fit of Ref. \cite{mrst}. Note,
however, that in Ref. \cite{mrst} a different prescription \cite{mrstcharm}
for the charm quark treatment, instead of the FFS, was
used\footnote{Actually, in Ref. \cite{mrst} a variant of the ACOT scheme
\cite{acot} was implemented. See Ref. \cite{coll} for an extensive review
on this issue.} and the value $m_c = 1.35\, GeV$. Fig.s 3-13 show the
comparison with experimental data of the structure functions calculated for
the values of parameters of Fit 1 (solid lines). We get a very good
description of polarized data, which, however, have larger errors than the
unpolarized ones. As far as the latter are concerned, a similar good
quality is exhibited by the neutrino structure function $x~ F_3$, while the
description of $F_2^\nu$ is not so accurate at low $x$. To some extent this
can be observed for the high-$x$ description of $F_2^p$ and $F_2^d$ too.
Note, however, that if the CCFR data are not included in the analysis, the
high-$x$ behaviour of $F_2^p$ and $F_2^d$ drastically improves
($\chi_{red}^2 = 0.97$ for the values of parameters reported in Table 2 as
Fit 1a), as is evident from the comparison with data of the dashed lines in
the figures. On one side, this suggests that, even with our hypothesis on
strange quarks, the discrepancy between NMC neutral current data and CCFR
charge current ones remains unresolved, and one has to consider other
effects like charge asymmetry \cite{bro96,boloth}. On the other side, the
high-$x$ difference between solid and dashed lines in Fig. 3 and 4 is also
justified by the fact that if the experimental ratio between $F_2^p$ and
$x~F_3$ at high-$x$ is different from 4/9 (the value implied by the
dominance of $u^\upa$ in that region) one cannot describe in a fully
satisfactory way both the structure functions unless allowing more freedom
in the quark parametrizations. 

With respect to the results found in our previous LO analysis \cite{posi}
we can confirm the pattern of the ratios $p_i^{(2)}/p_i^{(1)}$ between the
second and the first moments of the {\it gas} component of the
distributions of $u^\upa$, $u^\dna$, $d^\upa$, and $d^\dna$ quarks, and we
get similar values for the {\it temperature}, $\bar x$. Moreover, in
agreement with the conclusions by Brodsky and Ma \cite{bro96}, we find a
broader $s$ distribution than $\sbar$, as shown by the higher value of the
second moment of $s$ with respect to $\sbar$. More precisely, we have for
Fit 1 (similar values result for Fit 2 and 3) 
\be
p^{(2)} + \frac{2}{4.2}~ p_L^{(2)} = \left\{ 
\ba{ll}
0.0421 & {\rm for}\quad s, \\
0.0395 & {\rm for}\quad \sbar,
\ea
\right.
\ee
($p_L^{(2)}$ is the second moment of the {\it liquid} component) with a
relative difference of $\sim 7 \%$, compared to the value of $10\%$
obtained in \cite{bro96}. 

Finally, we can conclude that the fact that with the present NLO analysis
we find $\chi_{red}^2 < 2$ (and $<1$ for Fit 1a), which is smaller than the
value found in our previous approach \cite{posi}, is a good point in favour
of statistical distributions. 

\bigskip

\section*{Acknowledgments}

We thank Bo-Qiang Ma for stimulating our interest in studying the strange
parton distribution and for valuable and interesting observations.

\begin{table}[p]
\begin{center}
\begin{small}
\begin{tabular}{||c||c|c||}
\hline\ru1 
& $\al$ & $\bet$ \\ 
\hline\ru1 
$F_\nmc$ & -0.078 & 0.077 \\
\hline\ru1 
$F_\ccfr$ & -0.99 & 0.500 \\
\hline\ru1 
$F_c^\nc$ & -0.096 & 1.210 \\
\hline\ru1 
$F_c^\cc$ & -0.096 & 1.210 \\
\hline\ru1 
$F_1$ & -0.503 & 0.006 \\
\hline\ru1 
$g_1$ & 1.084 & 3.550 \\
\hline
\end{tabular}
\end{small}
\end{center}
\label{t:albet} 
\caption{Values of $\al$ and $\bet$ in the Jacobi description of the
structure functions relative to the NMC and CCFR experiments, to the charm
component in neutral and charge current processes, and to $F_1$ and $g_1$,
respectively.} 
\end{table}

\begin{table}[p]
\begin{center}
\begin{small}
\begin{tabular}{||c||c|c|c||c|c|c||}
\hline
\ru1 & \multicolumn{3}{c||}{Fit 1} & \multicolumn{3}{c|}{Fit 1a} \\
\hline\ru1 
$\chi_{red}^2$ & \multicolumn{3}{c||}{1.76} & \multicolumn{3}{c|}{0.97} \\
\hline\ru1 
$f_q(x)$ & \multicolumn{3}{c||}{$3.55~ x^{-0.275}~ (1-x)^{2.76}$} & 
\multicolumn{3}{c|}{$3.34~ x^{-0.254}~ (1-x)^{2.62}$} \\
\hline\ru1 
$f\qbar(x)$ & \multicolumn{3}{c||}{$7.33~ x^{-0.275}~ (1-x)^{4.84}$} & 
\multicolumn{3}{c|}{$8.28~ x^{-0.254}~ (1-x)^{11.2}$} \\
\hline\ru1 
$f_g(x)$ & \multicolumn{3}{c||}{$87.4~ x^{-0.276}~ (1-x)^{7.36}$} & 
\multicolumn{3}{c|}{$88.0~ x^{-0.255}~ (1-x)^{8.57}$} \\
\hline\ru1 
$f_L(x)$ & \multicolumn{3}{c||}{$0.124~ x^{-1.14}~ (1-x)^{14.7}$} & 
\multicolumn{3}{c|}{$0.120~ x^{-1.15}~ (1-x)^{12.6}$} \\
\hline\ru1 
$\bar x$ & \multicolumn{3}{c||}{0.250} & \multicolumn{3}{c|}{0.257} \\
\hline\ru1 
$\dde G$ & \multicolumn{3}{c||}{-0.009} & \multicolumn{3}{c|}{-0.193} \\
\hline\ru1 
& $\tilde x_i$ & $p_i^{(1)}$ & $p_i^{(2)}/p_i^{(1)}$ & $\tilde x_i$ & 
$p_i^{(1)}$ & $p_i^{(2)}/p_i^{(1)}$ \\ 
\hline\ru1 
$u^\upa$ & 0.750 & 1.58 & 0.149 & 0.867 & 1.50 & 0.161 \\
\hline\ru1 
$d^\dna$ & 0.150 & 0.870 & 0.118 & 0.184 & 0.836 & 0.126 \\
\hline\ru1 
$u^\dna$ & -0.033 & 0.581 & 0.109 & 0.047 & 0.639 & 0.119 \\
\hline\ru1 
$d^\upa$ & -0.136 & 0.440 & 0.105 & -0.100 & 0.447 & 0.113 \\
\hline\ru1 
$g^\upa$ & -0.427 & 3.88 & 0.055 & -0.409 & 3.93 & 0.052 \\
\hline\ru1 
$g^\dna$ & -0.426 & 3.89 & 0.055 & -0.399 & 4.12 & 0.052 \\
\hline\ru1 
& $k_i$ & & & $k_i$ & & \\
\hline\ru1 
$\ubar$ & 0.099 & 0.181 & 0.073 & 0.118 & 0.155 & 0.045 \\
\hline\ru1 
$\dbar$ & 0.181 & 0.330 & 0.073 & 0.232 & 0.303 & 0.045 \\
\hline\ru1 
$s$ & 0.114 & 0.122 & 0.095 & 0.112 & 0.109 & 0.101 \\
\hline\ru1 
$\sbar$ & - & 0.122 & 0.073 & - & 0.109 & 0.045 \\
\hline
\end{tabular}
\end{small}
\end{center}
\label{t:fitpar} 
\caption{Values of the parameters of the input distributions for Fit 1 and
Fit 1a (see text). We denote the first and second moments of the {\it gas}
component of the distributions with $p_i^{(1)}$ and $p_i^{(2)}$,
respectively.} 
\end{table}

\begin{table}[p]
\begin{center}
\begin{small}
\begin{tabular}{||c||c|c|c||c|c|c||c|c|c||}
\hline
\ru1 & \multicolumn{3}{c||}{Fit 1} & \multicolumn{3}{c||}{Fit 2} &
\multicolumn{3}{c||}{Fit 3} \\
\hline\ru1 
& $\dde k_i$ & $p_i^{(1)}$ & $p_i^{(2)}$ & $\dde k_i$ & $p_i^{(1)}$ &
$p_i^{(2)}$ & $\dde k_i$ & $p_i^{(1)}$ & $p_i^{(2)}$ \\
\hline\ru1 
$\dde \ubar$ & -1.00 & -0.181 & -0.013 & -1.00 & -0.185 & -0.013 & - & - &
- \\
\hline\ru1 
$\dde \dbar$ & 0.24 & 0.080 & 0.006 & 0.50 & 0.017 & 0.001 & - & - & - \\
\hline\ru1 
$\dde s + \dde \sbar$ & -1.00 & -0.243 & -0.020 & - & - & - & -1.00 &
-0.208 & -0.018 \\
\hline\ru1 
$a_8$ & \multicolumn{3}{c||}{0.956} & \multicolumn{3}{c||}{0.373} &
\multicolumn{3}{c||}{0.884}\\
\hline\ru1 
Bj & \multicolumn{3}{c||}{0.174} & \multicolumn{3}{c||}{0.175} &
\multicolumn{3}{c||}{0.176}\\
\hline\ru1 
EJp & \multicolumn{3}{c||}{0.133} & \multicolumn{3}{c||}{0.134} &
\multicolumn{3}{c||}{0.136}\\
\hline\ru1 
EJn & \multicolumn{3}{c||}{-0.041} & \multicolumn{3}{c||}{-0.041} &
\multicolumn{3}{c||}{-0.040}\\
\hline
\end{tabular}
\end{small}
\end{center}
\label{t:polpar}
\caption{Values of the polarization parameters and sum rules for the three
cases considered in the analysis.} 
\end{table}

\newpage
\pagestyle{empty}

\begin{figure}[p]
\begin{center}
\epsfig{file=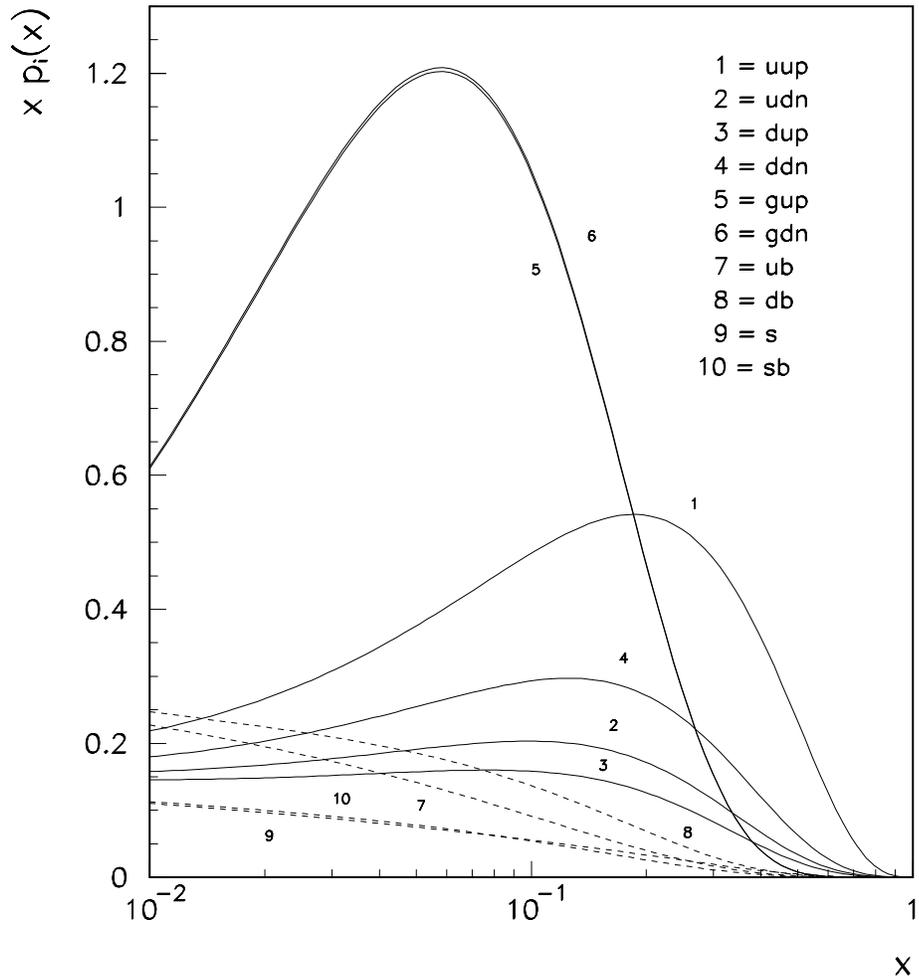,height=14truecm}\quad
\end{center}
\bigskip
\caption{Quark and gluon distributions at $Q_0^2 = 3\, GeV^2$ for the
values of parameters of Fit 1.} 
\label{f:distr}
\end{figure}

\newpage

\begin{figure}[p]
\begin{center}
\epsfig{file=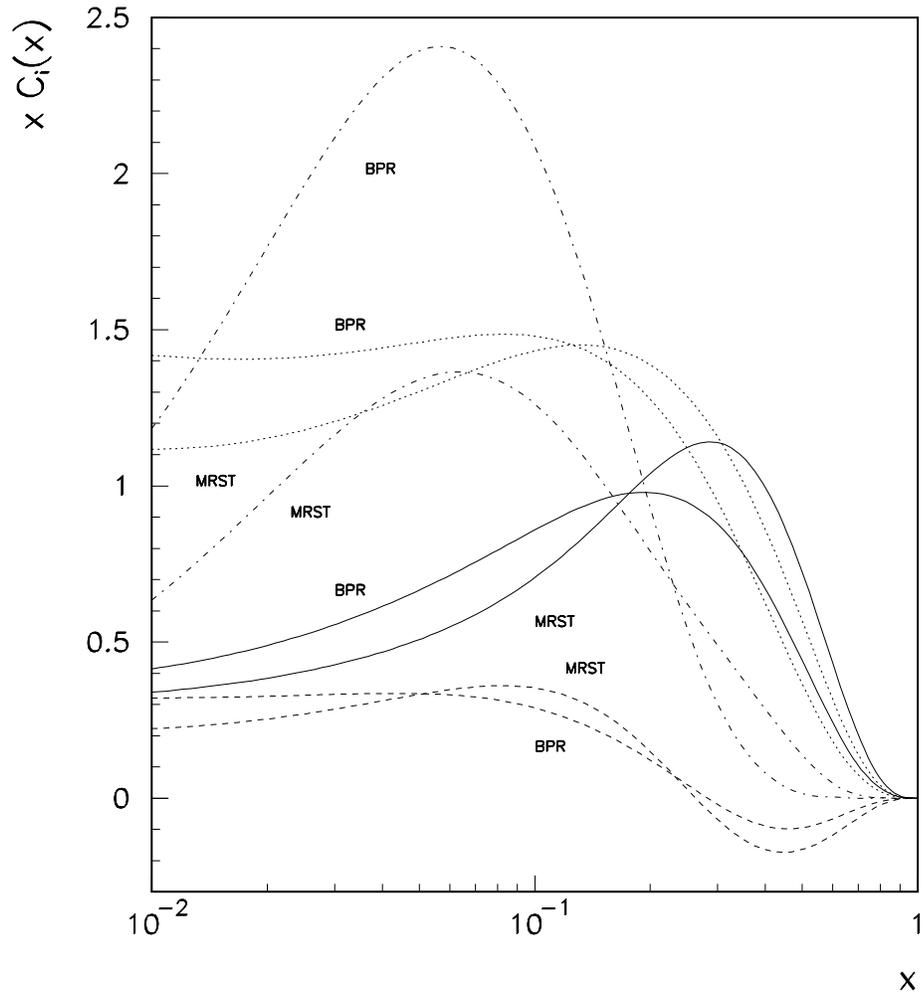,height=14truecm}\quad
\end{center}
\bigskip
\caption{Comparison of $Q_p$ (solid lines), $Q_n$ (dashed lines), $\Sigma$
(dotted lines), and $g$ (dash-dotted lines), evolved to $Q^2 = 1\,GeV^2$
(BPR), with the same quantities obtained from the fit of Ref.
\protect\cite{mrst} (MRST).} 
\label{f:comp}
\end{figure}

\newpage

\begin{figure}[p]
\begin{center}
\epsfig{file=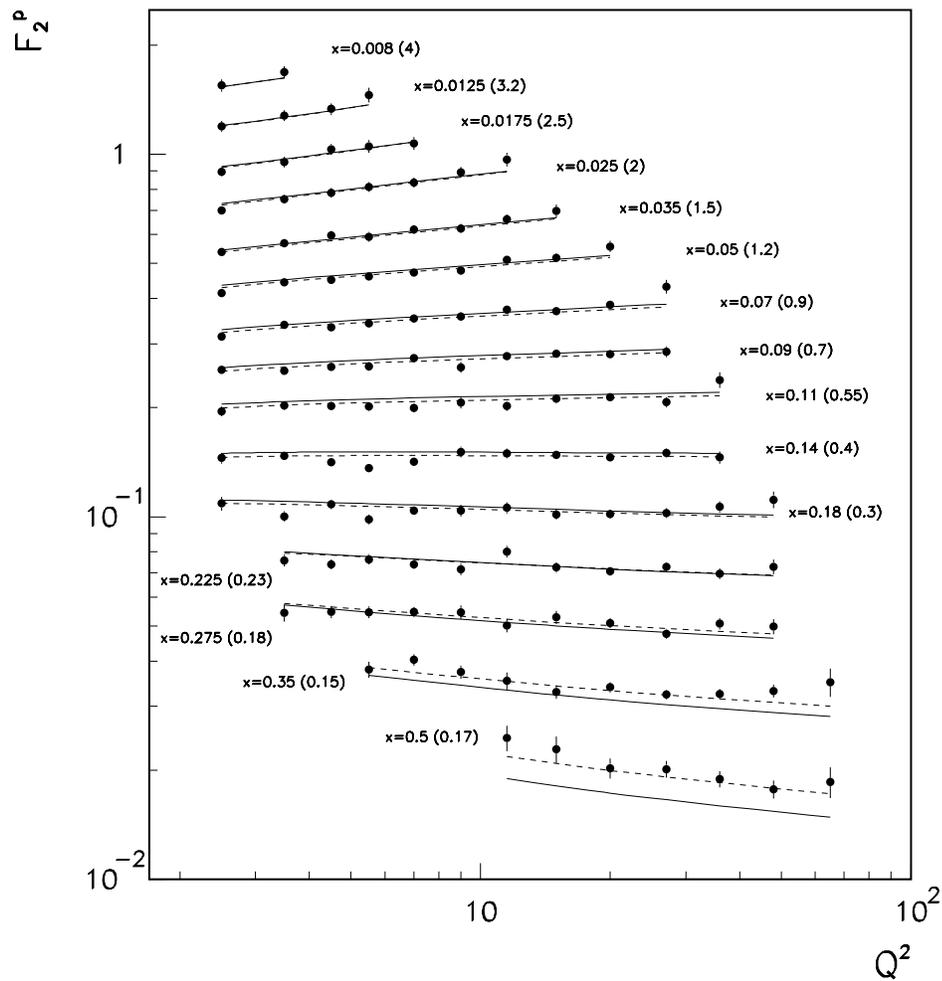,height=14truecm}\quad
\end{center}
\bigskip
\caption{Comparison of the prediction of the fit with the experimental data
on $F_2^p$ from NMC \protect\cite{nmcnew}. For display purposes $F_2^p$ has
been multiplied by the numbers in brackets. The solid and dashed lines
correspond to the value of parameters for Fit 1 and 1a, respectively (see
text). This notation is hereafter adopted.} 
\label{f:f2p}
\end{figure}

\newpage

\begin{figure}[p]
\begin{center}
\epsfig{file=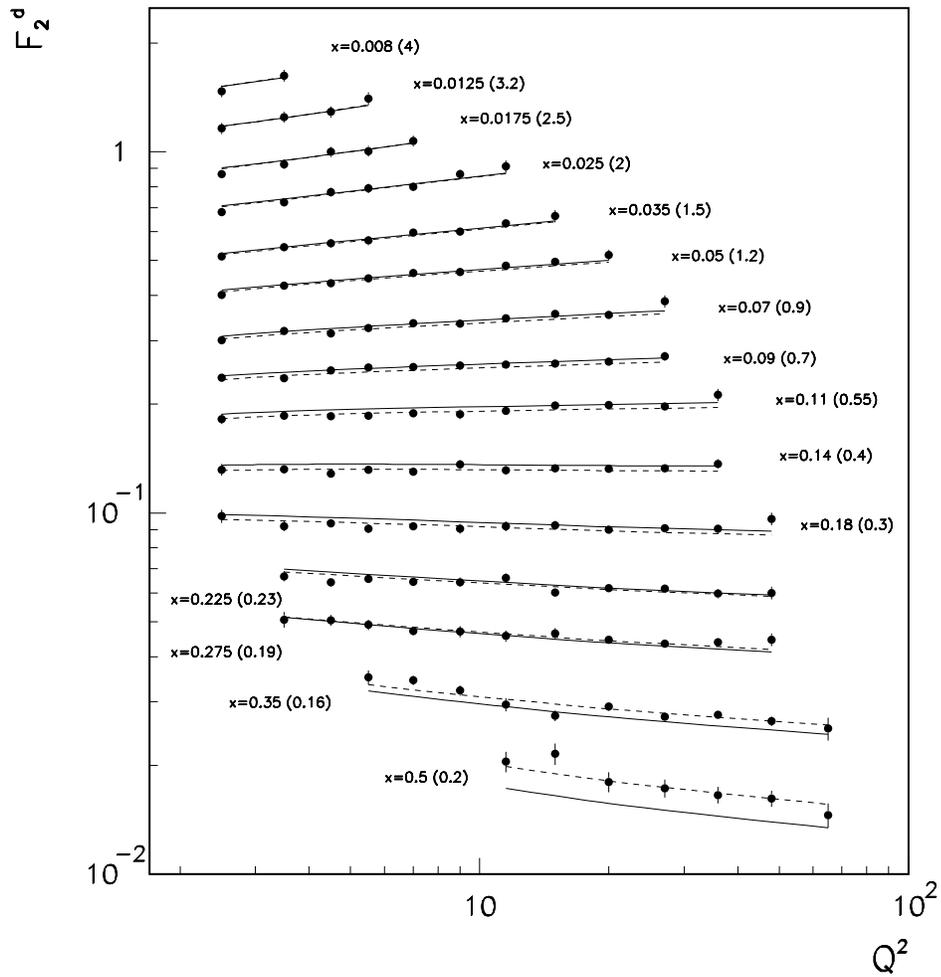,height=14truecm}\quad
\end{center}
\caption{Comparison of the prediction of the fit with the experimental data
on $F_2^d$ from NMC \protect\cite{nmcnew}. For display purposes $F_2^d$ has
been multiplied by the numbers in brackets.} 
\label{f:f2d}
\end{figure}

\begin{figure}[p]
\begin{center}
\epsfig{file=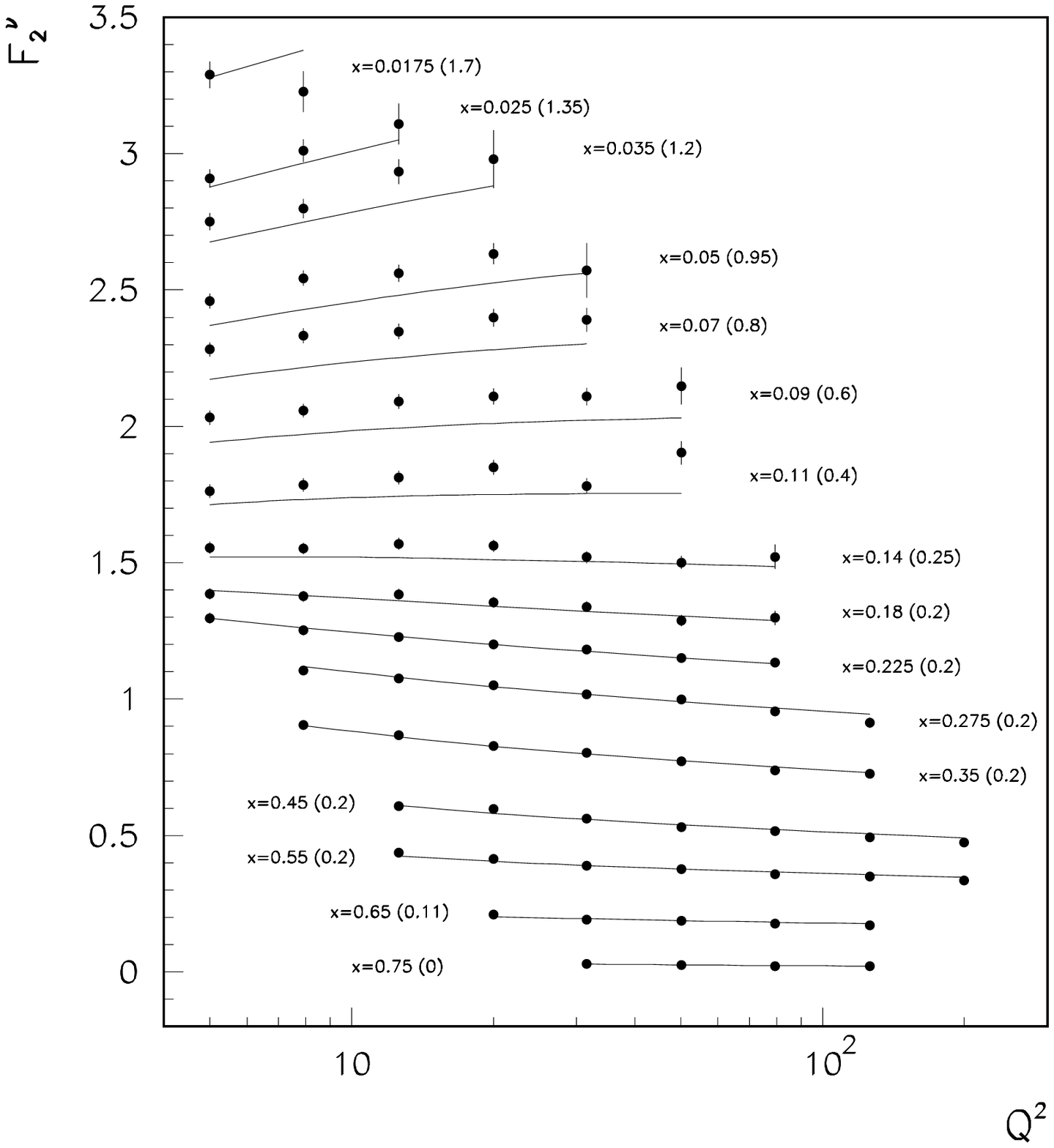,height=14truecm}\quad
\end{center}
\caption{Comparison of the prediction of the fit with the experimental data
on $F_2^\nu$ from CCFR \protect\cite{ccfrnew}. For display purposes the
numbers in brackets have been added to $F_2^\nu$.} 
\label{f:f2nu}
\end{figure}

\begin{figure}[p]
\begin{center}
\epsfig{file=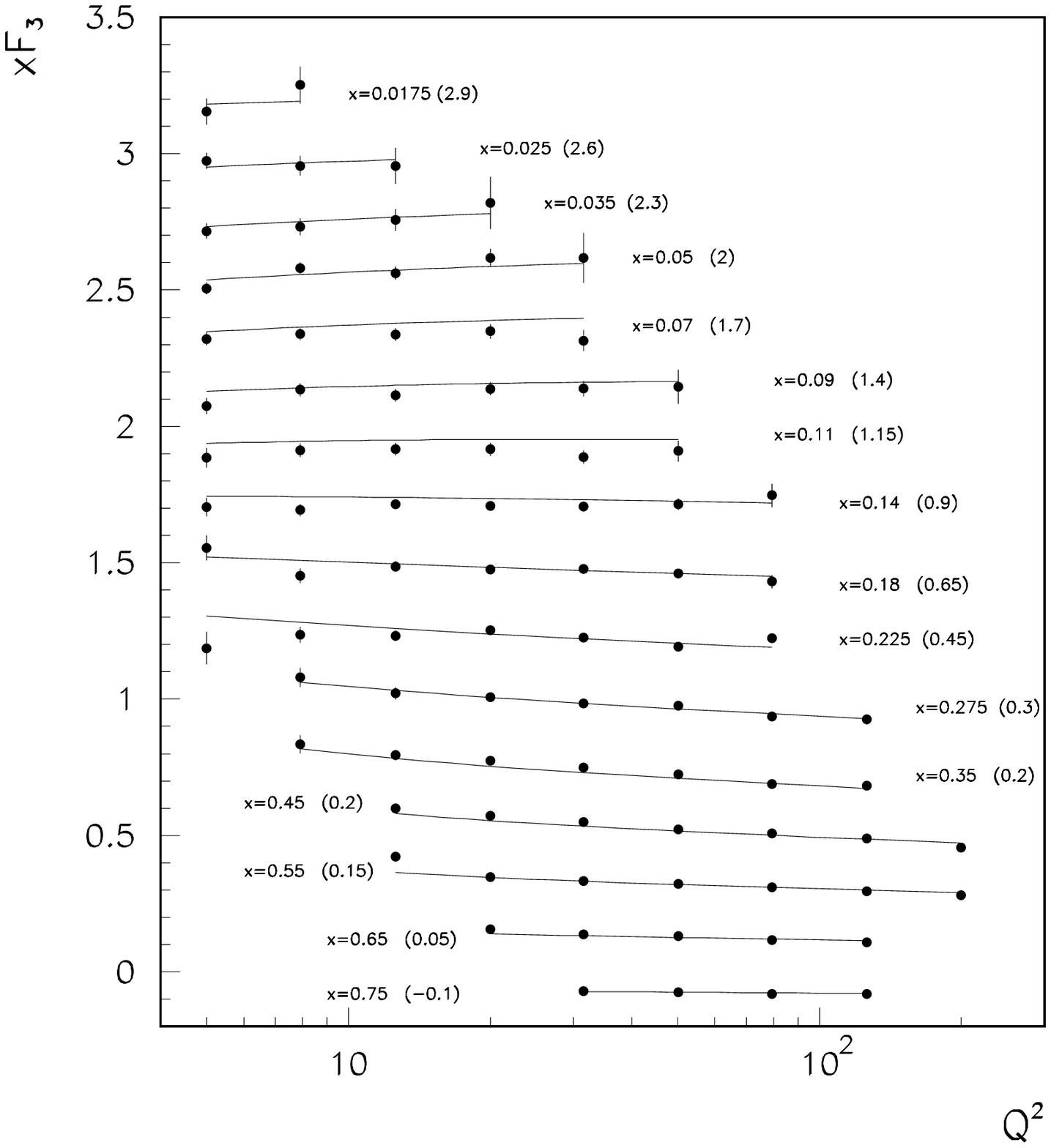,height=14truecm}\quad
\end{center}
\caption{Comparison of the prediction of the fit with the experimental data
on $x~F_3$ from CCFR \protect\cite{ccfrnew}. For display purposes the
numbers in brackets have been added to $x~F_3$.} 
\label{f:xf3}
\end{figure}

\begin{figure}[p]
\begin{center}
\epsfig{file=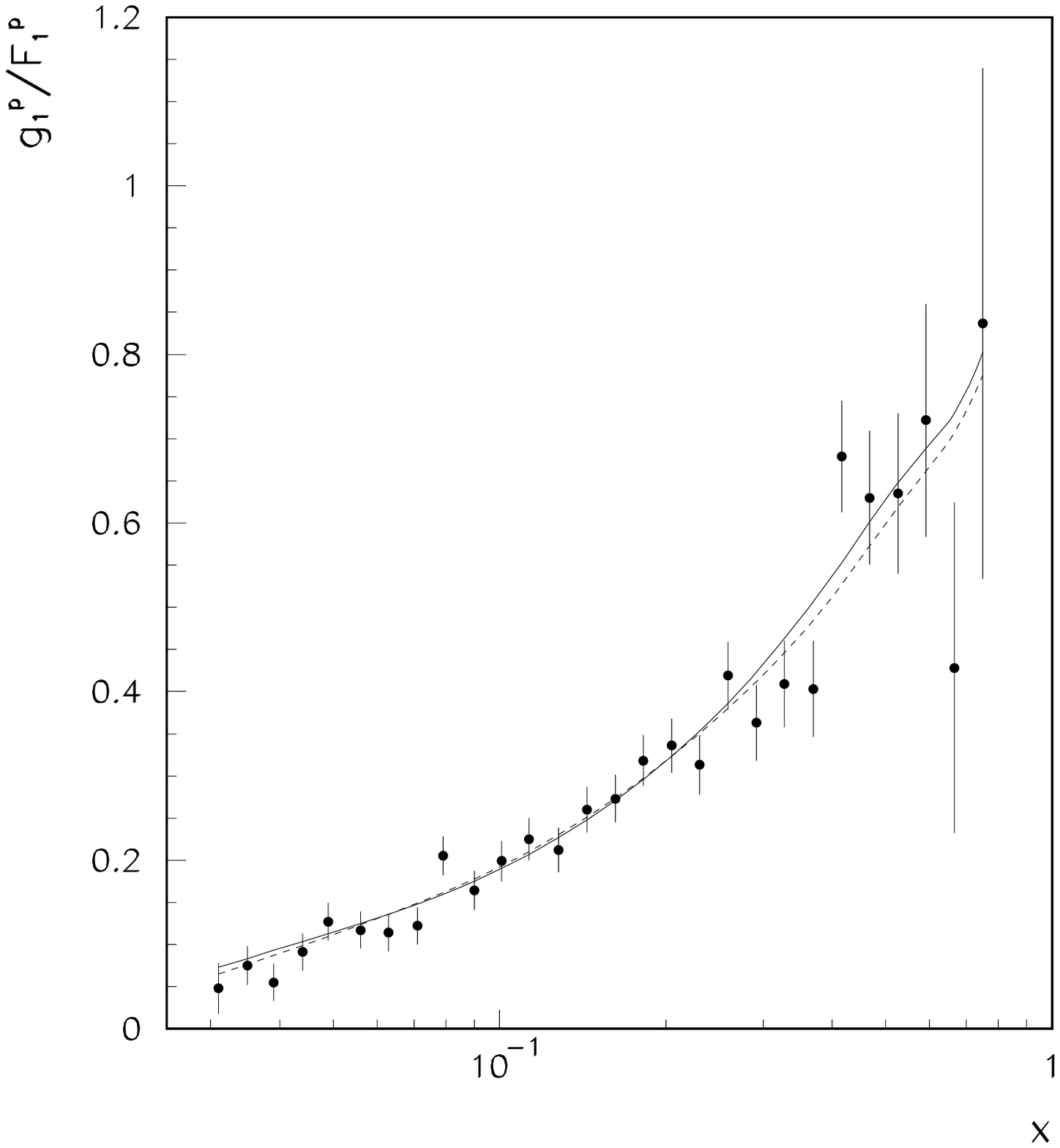,height=14truecm}\quad
\end{center}
\caption{Comparison of the prediction of the fit with the experimental data
on $g_1^p/F_1^p$ from E143 \protect\cite{e143}. The lines are evaluated at
the $Q^2$ of the experimental points.} 
\label{f:e143p}
\end{figure}

\begin{figure}[p]
\begin{center}
\epsfig{file=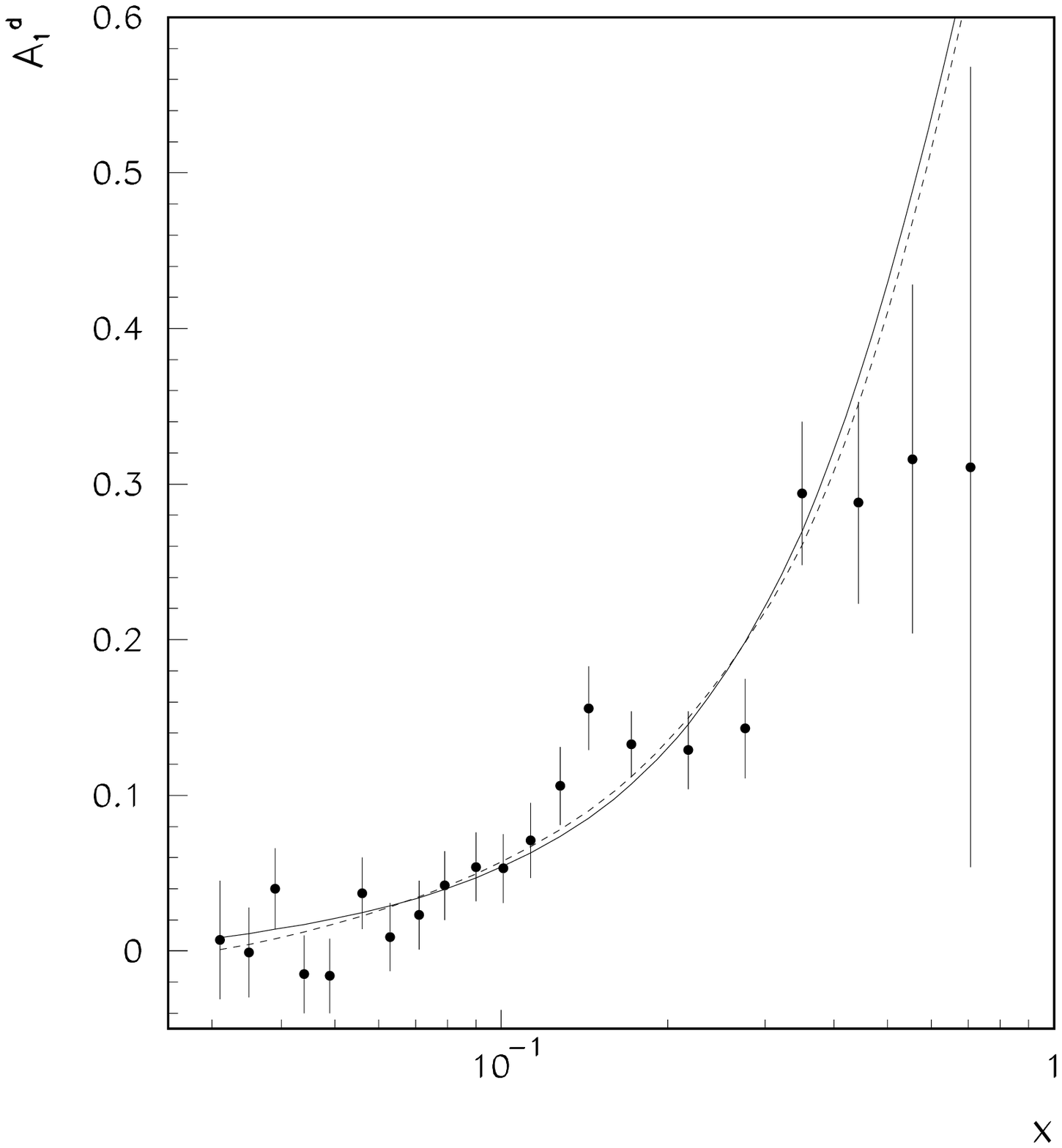,height=14truecm}\quad
\end{center}
\caption{Comparison of the prediction of the fit with the experimental data
on $A_1^d$ from E143 \protect\cite{e143}. The lines are evaluated at the
$Q^2$ of the experimental points.} 
\label{f:e143d}
\end{figure}

\begin{figure}[p]
\begin{center}
\epsfig{file=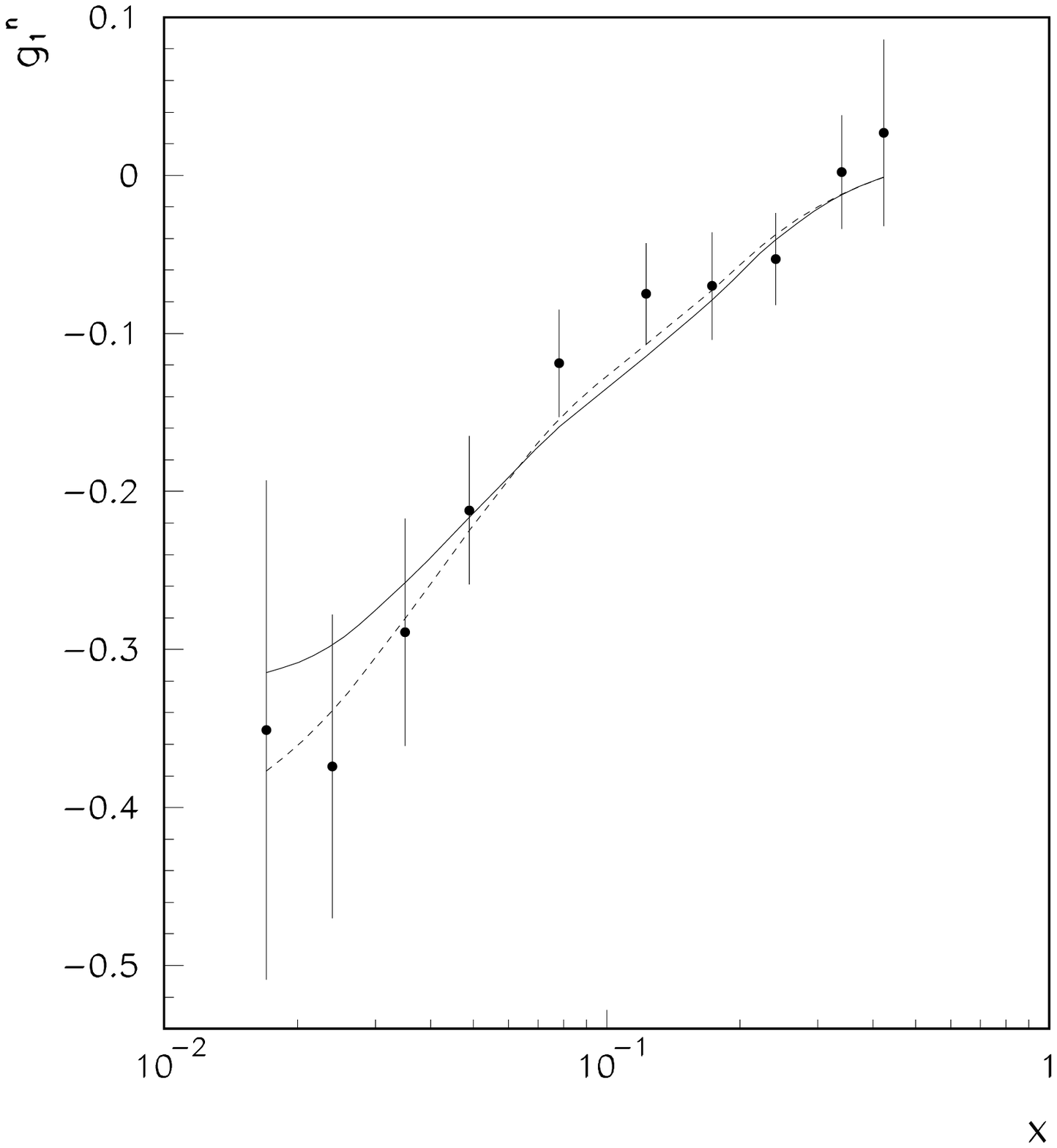,height=14truecm}\quad
\end{center}
\caption{Comparison of the prediction of the fit with the experimental data
on $g_1^n$ from E154 \protect\cite{e154}. The lines are evaluated at the
$Q^2$ of the experimental points.} 
\label{f:e154g}
\end{figure}

\begin{figure}[p]
\begin{center}
\epsfig{file=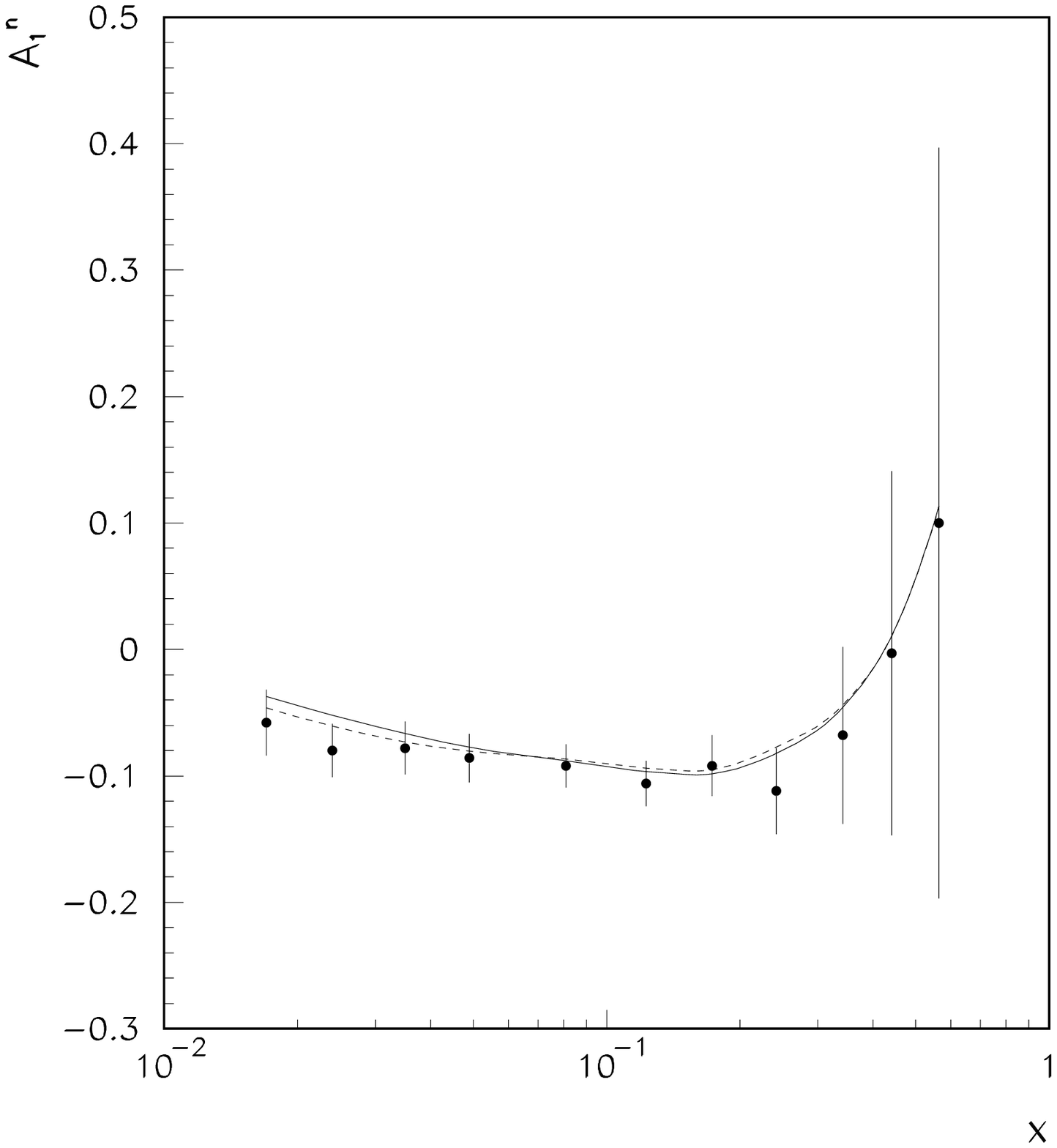,height=14truecm}\quad
\end{center}
\caption{Comparison of the prediction of the fit with the experimental data
on $A_1^n$ from E154 \protect\cite{e154}. The lines are evaluated at the
$Q^2$ of the experimental points.} 
\label{f:e154a}
\end{figure}

\begin{figure}[p]
\begin{center}
\epsfig{file=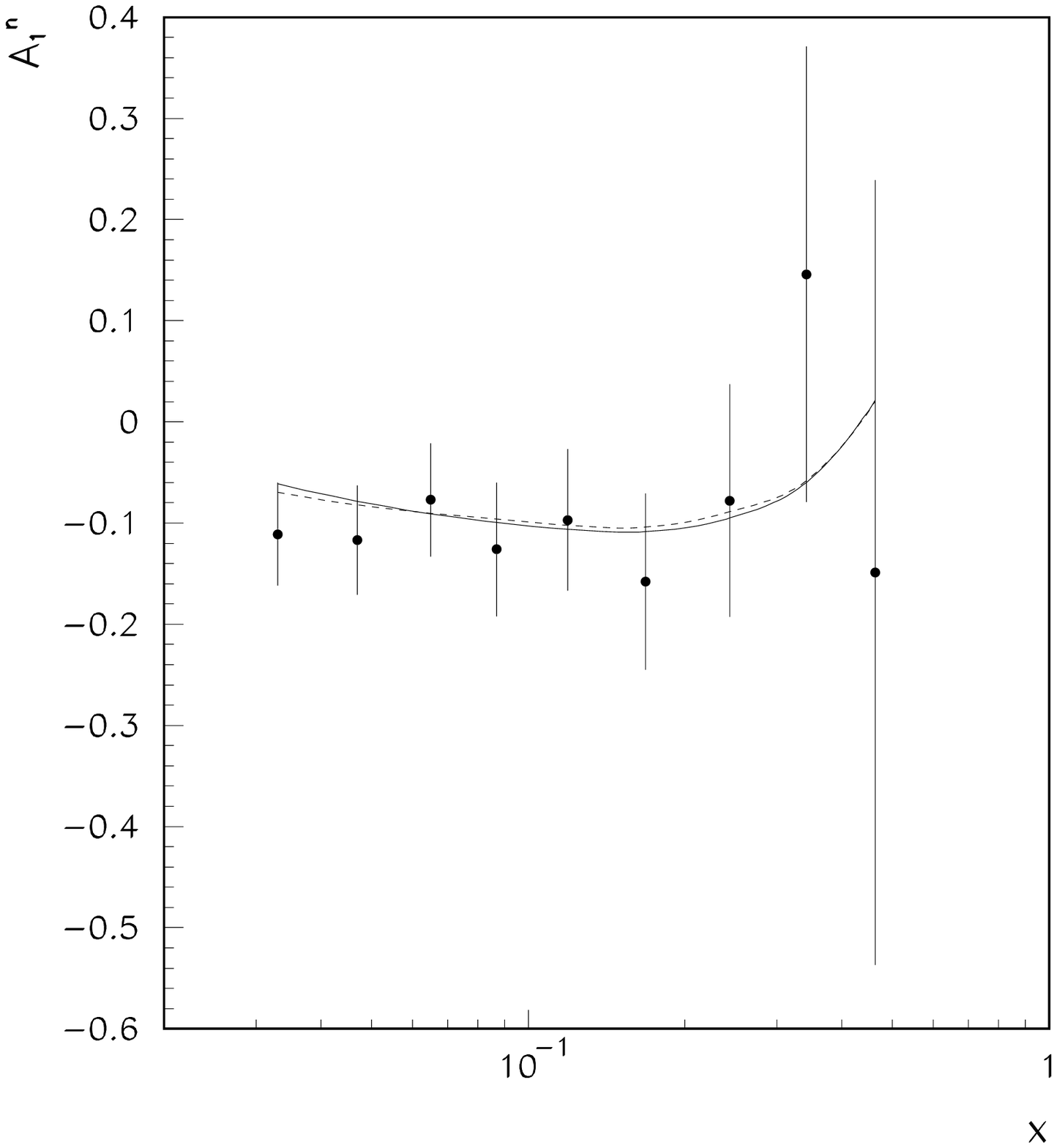,height=14truecm}\quad
\end{center}
\caption{Comparison of the prediction of the fit with the experimental data
on $A_1^n$ from HERMES \protect\cite{hermes}. The lines are evaluated at
the $Q^2$ of the experimental points.} 
\label{f:hermes}
\end{figure}

\begin{figure}[p]
\begin{center}
\epsfig{file=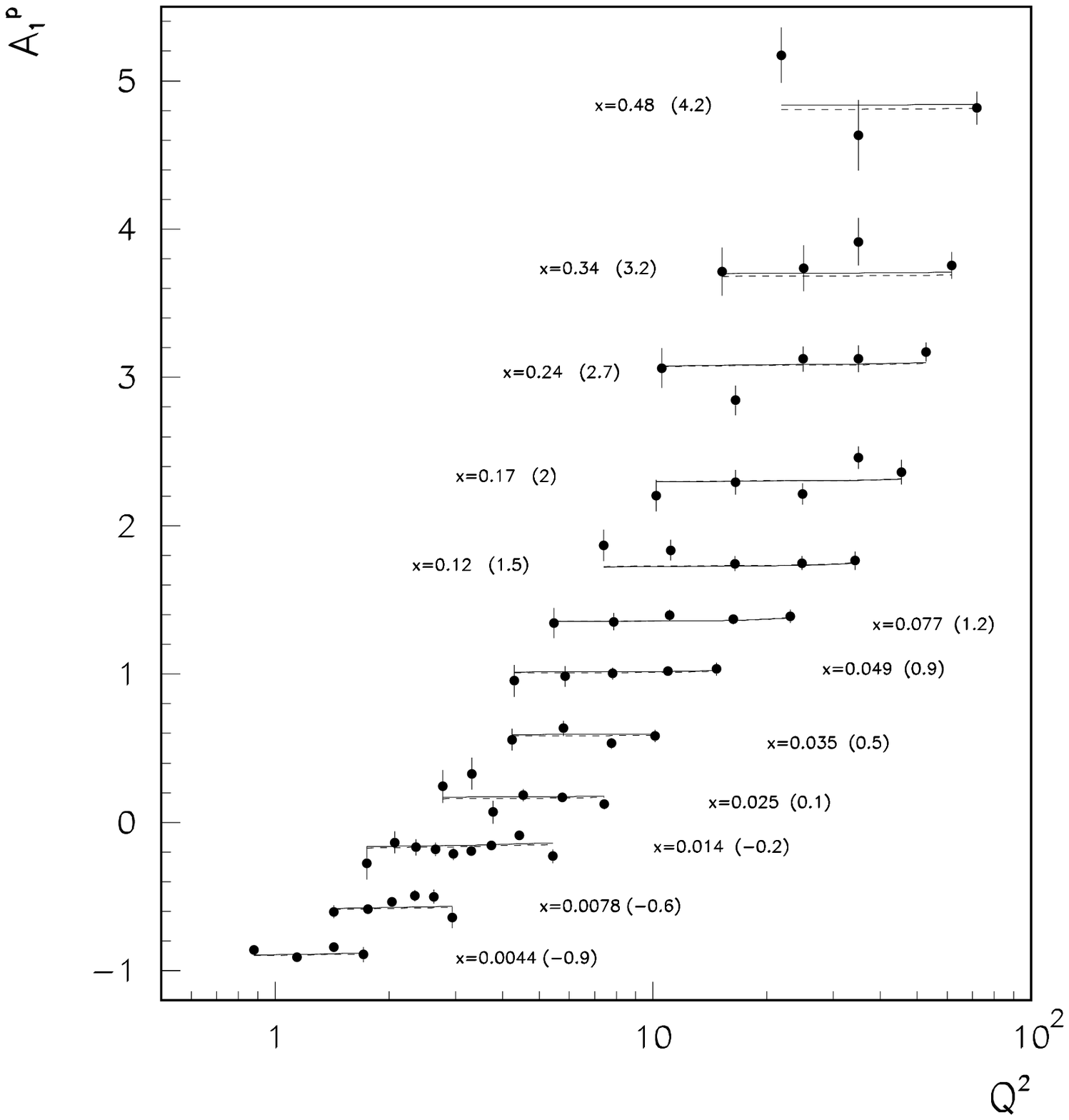,height=14truecm}\quad
\end{center}
\caption{Comparison of the prediction of the fit with the experimental data
on $A_1^p$ from SMC \protect\cite{smc}. For display purposes data at
adjacent $x$ values have been grouped together and the numbers in brackets
have been added to $A_1^p$.} 
\label{f:smcp}
\end{figure}

\begin{figure}[p]
\begin{center}
\epsfig{file=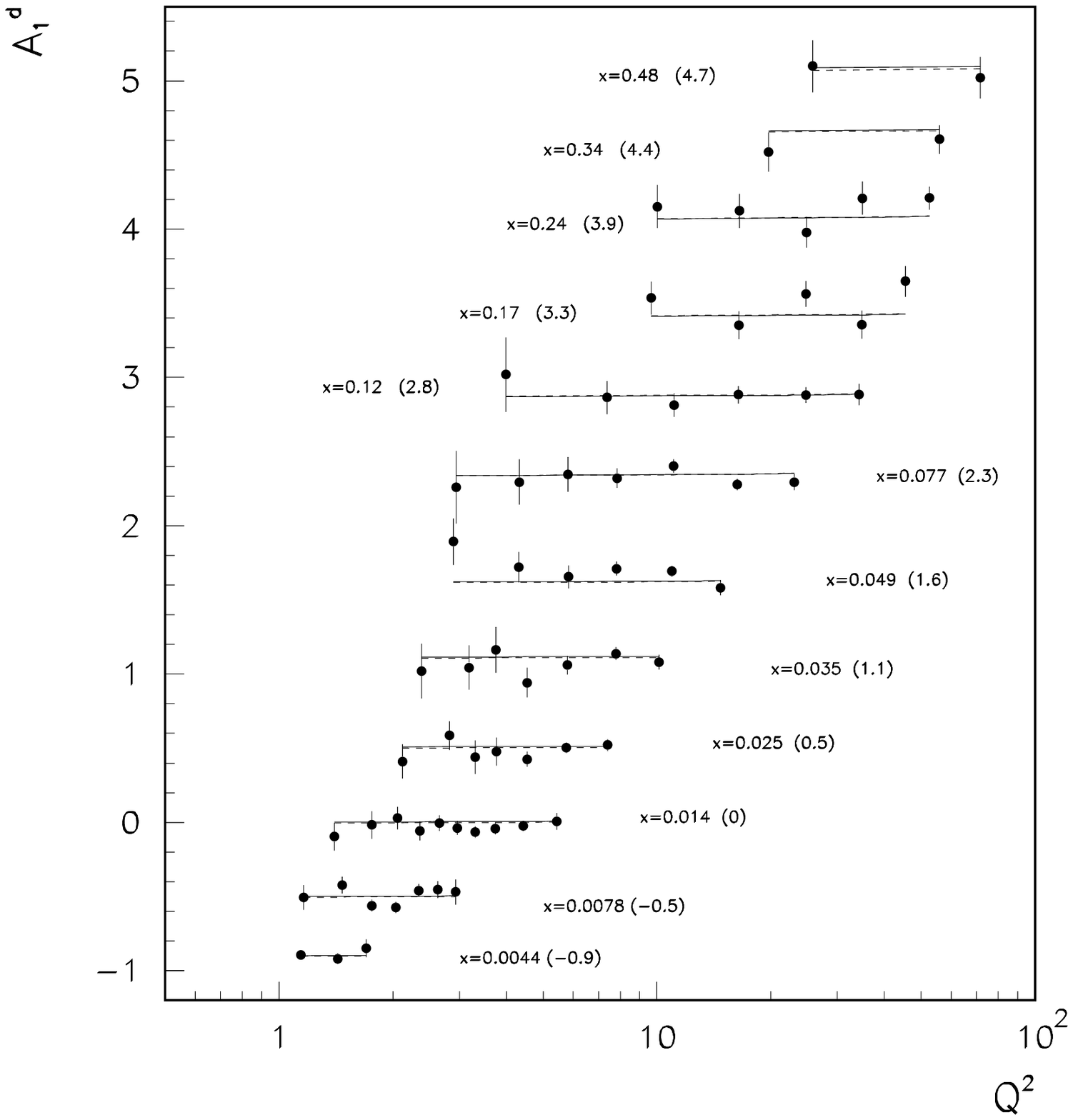,height=14truecm}\quad
\end{center}
\caption{Comparison of the prediction of the fit with the experimental data
on $A_1^d$ from SMC \protect\cite{smc}. For display purposes data at
adjacent $x$ values have been grouped together and the numbers in brackets
have been added to $A_1^d$.} 
\label{f:smcd}
\end{figure}

\end{document}